\documentclass[aps,pre,twocolumn,showpacs,groupedaddress]{revtex4-1}

\usepackage{amssymb,amsmath,graphicx,bm,textcomp,xcolor,commath,bm}
\usepackage{ulem}
\usepackage[utf8]{inputenc}

\def \bft {{\bf t}}

\def \bfe {{\bf e}}
\def \LL {\mathcal{L}}
\def \HT {\mathcal{H}}

\begin{document}

\title{"Twist-Controlled" force amplification \& Spinning tension transition in yarn}

\author{Antoine Seguin}\affiliation{Universit\'e Paris-Saclay, CNRS, Laboratoire FAST, F-91405, Orsay, France}
\author{J\'er\^ome Crassous} \affiliation{Univ Rennes, CNRS, IPR (Institut de Physique de Rennes) - UMR 6251, F-35000 Rennes, France}
\email{jerome.crassous@univ-rennes1.fr}

\date{\today}

\begin{abstract}
Combining experiments and numerical simulations with a mechanical/statistical model of twisted yarns, we discuss the spinning transition between a cohesion-less assembly of fibers into a yarn.  We show that this transition is continuous but very sharp due to a giant amplification of frictional forces which scales as $\exp \theta ^2$, where $\theta$ is the twist angle. We demonstrate that this transition is controlled solely by a non-dimensional number $\HT$ involving twist, friction coefficient, and geometric lengths. A critical value of this number $\HT_c \simeq 30$ can be linked to a locking of the fibers together as the tensile strength is reached. This critical value imposes that yarns must be very slender structures with a given pitch. It also induces the existence of an optimal yarn radius. Predictions of our theory are successfully compared to yarns made from natural cotton fibers.
\end{abstract}

\maketitle

Yarns made from natural fibers are one of the first materials ever processed by humans, including Neanderthals~\cite{hardy.2020}. They are done by making bundles of initially aligned fibers which are then stuck together by twisting. The fact that many individual fibers a few centimeters may form yarns of tens of meters drew early attention from scientists. Galileo~\cite{galileo} argued that the twist "binds" the filaments together, but do not discuss the origin of this cohesion. We now know that the binding forces are created by the tension throughout the filaments which creates normal forces due to the curvatures of the fibers, and that tangential frictional forces prevents sliding of fibers~\cite{birebent.1929,bouasse,barella.1962}. If the twist is large enough, the relative sliding of fibers are totally blocked, and the rupture of the yarn is then a problem of statistic of rupture of individual fibers~\cite{peirce.1926,pradhan.2010}. The description of the transition between fibers which are "free to slide" without spinning, to "blocked by spinning" is still a open problem. Experimentally, only very few studies addressed the dependence of yarn strength with twist level~\cite{gegauff1907}. Theoretically, despite numerous attempts, the mechanism linking twist and strengthening has not been clearly understood~\cite{sullivan.1942,hearle.1965,broughton.1992,pan.1992,pan.1993}. Recently, an analogy with the percolation transition had been suggested~\cite{warren.2018}. Assembly of fibers is an example of assembly of objects that interact through numerous frictional contacts. For such systems, the geometrical arrangement of the contact points may generate huge stress throughout the system. Some examples of such systems are granular materials in proximity to a solid wall (Jansen effect~\cite{andreotti.book,Bertho2003}), assembly of parallel sheets in contact (Interleaved phone book experiment~\cite{alarcon.2016,taub.2021}), or contact points distributed around a cylinder (Capstan). In all those examples, the proportionality between the tangential and the normal stress at contact means that the mechanical stress in the system decreases exponentially with the distance to applied load, and then have drastic effects of the mechanical equilibrium of such system.

\begin{figure}[t]
\centering
\includegraphics[width=\columnwidth]{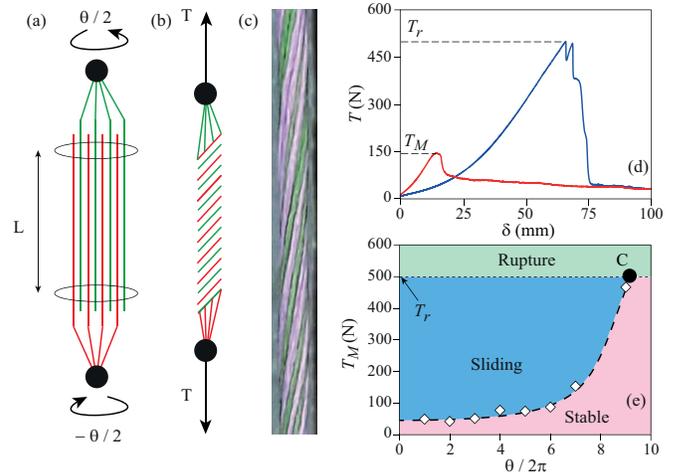}
\caption{(a,b) Preparation of the model yarn before (a) and after (b) twisting. (c) Photo of a yarn made from cotton strings after twisting. (d) Traction forces as function of displacement for cotton yarn $L=800$~mm: (blue) $\theta / 2\pi =11$, (red) $\theta / 2\pi=3$. (e) Symbol: Maximum traction force as full twist angles (cotton yarn, $L=800$~mm), dotted line is for eye guide, and dashed line is the rupture force.}
\label{fig1}
\end{figure}

\begin{figure*}[t]
\centering
\includegraphics[width=\linewidth]{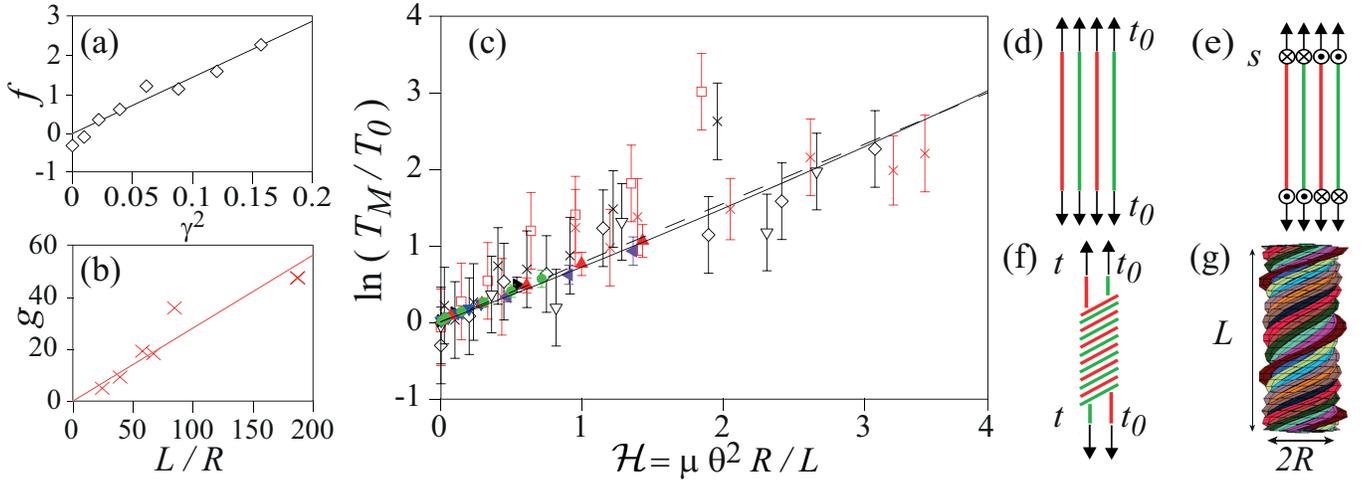}
\caption{
(a) Scaling law $f(\gamma^2)$ for cotton yarn at fixed $R$ and $L$. Line is linear fit. (b) Scaling law $g(L/R)$ for flax yarn at fixed twist $\theta=2.5~turns$.  Line is linear fit. (c) $\ln(T_M/T_0)$ as function $\mathcal{H}$. Dashed line is eq.\eqref{eq2}, plain curve is eq.\eqref{eq6b}. For (a,b,c) Crosses and open symbols are experimental data. Cotton, $N=20$, $R=3.15$~mm: $L=200$~mm ($\triangledown$), $L=400$~mm ($\lozenge$), $L=200$~mm ($\triangledown$). Flax, $N=20$, $R=4.15~$mm: $L=400$~mm, various $\theta$ (\textcolor{red}{$\square$}), $\theta=2.5$~turns, various $L$ (\textcolor{red}{$\times$}). Plain symbols are numerical data with $L/R=60$: $\mu_m=1$, $N=40$ (\textcolor{red}{$\blacktriangle$}), $\mu_m=0.5$, $N=40$ (\textcolor{green}{$\bullet$}), $\mu_m=0.5$, $N=20$ (\textcolor{violet}{$\blacktriangleleft$}), $\mu_m=0.5$, $N=100$ ($\blacktriangleright$), $\mu_m=0.2$, $N=40$ (\textcolor{blue}{$\blacktriangledown$}). (d,e,f) Schematic drawing of the preparation of the numerical yarn: (d) uniform tension $t_0$ is applied; (e) shear force $s$ is applied to twist the yarn; (f) Tension is increased to $t$ on the top of half fibers, and on the bottom of the other fibers. (g) Snapshot of a brush of fibers after twisting, and during the increase of $t$ ($N=20$, $L/R=60$). Note the difference of vertical and horizontal scales.}
\label{fig2}
\end{figure*}

We show in this letter that an assembly of fibers belongs to the same class of system. For this we consider model yarns made of entangled twisted fibers. The tension necessary to unravel the fibers is shown to vary continuously, but very rapidly with the twist. This sharp evolution of the disentanglement force creates a phase transition like transition between free fibers and stuck fibers phases. A simple mechanical model of frictional helicoidal fibers allows us to define a non dimensional number whose value characterizes this transition. These results can be successfully applied to real yarns.

{\it Experimental model yarn system.} Our starting point is the demonstrating experiment of friction force in yarns as proposed by Bouasse~\cite{bouasse}. We consider two brushes of $N/2$ identical fibers (see fig.1(a)). The fibers are passed through rings which are connected to puller jaws ($N/2$ fibers in each jaw). The model fibers are of flexible strings of cotton (diameter $d=1$~mm, linear density $\lambda=0.48$~g.m$^{-1}$, friction coefficient $\mu_m=0.35$, bending modulus $B\sim 10^{-6}$~N.m$^2$), or flax ($d=1$~mm, $\lambda=1.03~g.m^{-1}$, $\mu_m=0.53$,  $B\sim 4.10^{-6}$~N.m$^2$). The twist of the elementary yarns composing each string are always very large compared to the twist that we apply. We first prepare the entanglement by alternately aligning the brushes roughly parallel. The brushes are then zipped together with two plastic cable clamps, and twisted by a angle $\theta$ (fig.1(b,c)). The puller jaws are attached to a traction measurement apparatus (Instron 5965, $5$~kN force sensor) and elongated at fixed velocity $50~$mm.min$^{-1}$. Fig.1(d) shows the force variations for two different twist angles. If the twist angle is low enough, the force first increases, reaches a peak value (noted $T_M$) and then decreases slowly. Such variations are associated to a smooth relative sliding of the two brushes. For large enough twist, a force drop is measured after the maximum force (noted $T_{r}$). This is associated to the rupture of one or many strings that we may observe by post-mortem inspection. The figure 1(e) shows the evolution of $T_M$ as a function of the twist angle. This value is likely constant up to $\theta / 2\pi \simeq 5$~revolutions for this yarn, and increases  rapidly up to $9~$revolutions where $T_M$ reaches $T_{r}$ at point C.

{\it Scaling laws for maximum traction.} We first limit our analysis to the maximum force $T_M$ and we do not discuss rupture. Since we expect that the maximum force is dependant of friction, $T_M$ should depend on $\mu_m$ and of geometric characteristics of the yarn : $\theta$, $L$, $R$ and $N$. We define the twist rate $\gamma=R\theta / L\ll 1$.
\indent We first discuss the $\gamma$ dependence of $T_M$. Noting $T_0$ the traction force at vanishing twist, we must have $T_M(\gamma)=T_0~F(\gamma)$, or $\ln(T_M)=\ln(T_0)+f(\gamma)$ with $f=\ln(F)$ an even function vanishing at $\gamma=0$. Leading term of expansion at small twist is $f\sim \gamma^2$. This dependence is experimentally verified as shown on fig.2(a). It follows that:
\begin{align}
\ln(T_M/T_0)=\gamma^2~g(L/R,N,\mu_m)
\end{align}
where $g$ is a non-dimensional function of non-dimensional parameters. The $L/R$ dependence of $g$ is obtained by considering the evolution of traction force at fixed $\theta$, $R$ and $N$ and of various lengths $L$. We found (see fig.2(b)) that $g(L/R,N,\mu_m) \sim L/R$, so that  $\ln(T_M/T_0)=(\gamma^2 L /R)~h(N,\mu_m)$.

{\it Numerical yarn.} We use discrete element method simulations~\cite{crassous.tbp} to obtain the function $h$. Fibers are modeled as set of point masses connected with elongational spring/dashpot without torsional or bending restoring forces. Successive masses are connected with cylinders of diameter $d$. The contact points between cylinders (belonging to same or different fibers) are calculated, and the contact forces are calculated considering normal stiffness and damping, and tangential stiffness with Coulomb friction coefficient $\mu_m$. Equations of motion are integrated using a Verlet algorithm. The steps for making numerical yarns is depicted in fig.2. We first stretch the $N$ fibers under a force $t_0$ (fig.2(d)) such that the strain of each fiber is $10^{-4}$. A torque is then applied
to the yarn by submitting both ends of fibers to orthoradial forces $s$ (fig.2.(e)). During this preloading phase, $\mu_m$ is kept to a low value $0.05$ which ensure an uniform twist along the yarn (fig.2(g)). Finally, while keeping forces $t_0$ and $s$ applied, the tension $t$ of half the fibers on the bottom and to the other half at the top (fig.2(f)) is slowly increased until a value $t=t_M$ where the brush separates.\\
\indent Full symbols of Fig.2(c) shows the evolution of $t_M/t_0$ with the twist angle for different values of $\mu_m$ and $N$. First, we obtain that $\ln{(t_M/t_0)}\sim \theta^2$ as for experimental data. We have also checked (data not shown here) that: $g\sim L/R$. The friction coefficient $\mu_m$ is varied, and the $N$-dependency is obtained from simulations of $N$ fibers of radius $a_N$ such that $R=a_N~\sqrt{N/\phi}$ (with $\phi=0.80$ the packing fraction) ensuring fixed string radius $R$. We did not identify significant variations with $N$ between $N=20$ and $N=100$ (fig.2(c)).

Finally, fig.2(c) shows that all the experimental and numerical data may be collapsed using the single law:
\begin{equation}
T_M/T_0=\exp{(0.75~\mu~\theta^2~\frac{R}{L})}\label{eq2}
\end{equation}
with $\mu= 0.63~\mu_m$ for laboratory and $\mu=1.13~\mu_m$ for numerical experiments. The experimental dependence on $\mu_m$ may be viewed on fig.2(c) where data for flax and cotton collapse when plotted as function of $\mu \theta^2 R/L$. Finally, the amplification of the tension in the yarn is thus exponential, and only related to a dimensionless number $\HT=\mu \theta^2 R / L$ that we name "Hercules twist Number".

\begin{figure}[t!]
\centering
\includegraphics[width=\columnwidth]{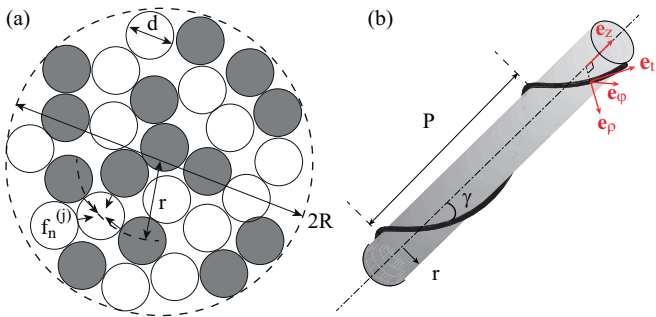}
\caption{(a) Section of a yarn of radius $R$ composed of fibers of diameter $d$. Gray fibers go downwards and white fibers go upwards. (b) A fiber twisted on a cylinder of radius $r$. }
\label{fig4}
\end{figure}

{\it Mechanical model.} We develop a mechanical model for deriving~\eqref{eq2}. We consider a yarn made of $N$ helicoidal fibers (fig.3(a)) with some rising and descending fibers. We consider first a twisted fiber at a distance $r$ from the axis: ${\bf r}=r {\bf e}_\rho+ z {\bf e}_z$ in cylindrical coordinates $(\rho,\varphi,z)$ (fig.3(b)). The geometry of the helix of constant pitch $P$ gives $\varphi / 2\pi= z/P$ and we define the reduced pitch as $p=P/2\pi$. For pitch large compared to $r$, the tangent vector of the fiber is ${\bf e}_{t}(z) \simeq  r / p \  {\bf e}_\varphi+{\bf e}_z$. The tension is $\bft(z)=t(z) {\bf e}_{t}(z)$, and:

\begin{align}
\frac{d\bft}{dz}=\frac{dt}{dz}\bfe_t-\frac{r}{p^2} t(z) \bfe_\rho\simeq\frac{dt}{dz}\bfe_z-\frac{r}{p^2} t(z) \bfe_\rho\label{eq3}
\end{align}

We first consider the force equilibrium, in a section of the yarn, for a portion of fiber between $z$ and $z+dz$. The force $- (r dz/p^2) t(z) \bfe_\rho $ is a linear restoring force towards the axis of the yarn: the torsion of the yarn is then equivalent putting the fiber into a twist-controlled harmonic potential $V(r)= t(z)dz~(r^2/2p^2)$. At mechanical equilibrium, contact forces must balance this confining force. The equilibrium of forces in the plane perpendicular to the fiber writes:
\begin{align}
\frac{r}{p^2} t(z) \bfe_\rho=\sum_{j=1}^{j=\mathcal{N}} f_n^{(j)} {\bf e}_n^{(j)}\label{eq4}
\end{align}
with ${\mathcal{N}}$ the number of contacts, $f_n^{(j)} dz~{\bf e}_n^{(j)}$ the contact force between $z$ and $z+dz$ exerted by fiber $j$, and ${\bf e}_n^{(j)}$ the normal vectors at contact points. Let $f_n$ be the order of magnitude of normal forces $f_n^{(j)}$. Since vectors ${\bf e}_n^{(j)}$ have random orientations, r.h.s. of \eqref{eq4} may be viewed as a $2d$ random walk in force space, and we should have: $t(z) r / p^2 \sim \sqrt{\mathcal{N}} f_n$. We now consider the force along $z$ of the rising fiber due to the $\mathcal{N}/2$ fibers that do not rise. Each contact exerts a sliding force $\simeq \mu_m f_n$, and then $(dt/dz) \simeq (\mathcal{N}/2) \mu_m f_n \simeq (\sqrt{\mathcal{N}}/2) \mu_m t(z) r/p^2$. We finally obtain:
\begin{align}
\frac{dt}{dz}=\mu \frac{r}{p^2} t(z)\label{eq5}
\end{align}
with $\mu=(\sqrt{\mathcal{N}}/2) \mu_m$. The coordination number for a random close packing of disks being $4$~\cite{vanhecke.2009}, we should have $\mu \simeq \mu_m$, in agreement with laboratory and numerical experiments. Integrating \eqref{eq5} along $z$ gives $t(L)=t_0~\exp{\left(\mu r L/p^2\right)}$. Using $\theta  = L / p$, and
 $dN(r)/dr=Nr/R^2$ the density of rising fibers, the force on the yarn section is:
\begin{subequations}
\begin{align}
T_M &= \int_{r=0}^{r=R} t_0~\exp{(\mu \theta^2 \frac{r}{L})}~dN(r)\label{eq6a}\\
&= T_0 \frac{2[(\mathcal{H}-1)\exp{\mathcal{H}}+1]}{\mathcal{H}^2}\label{eq6b}
\end{align}
\end{subequations}
where $T_0=Nt_0/2$, and with $\HT$ the Hercules twist Number $\mathcal{H}$ previously defined. Since $t_0$ is only in prefactor of the exponential amplification, the scaling $\ln(T_M/T_0) \sim \HT$ is expected to hold if \eqref{eq6a} is extended to a radius dependant tension $t_0(r)$, as it is the case for dense packing of twisted fibers~\cite{panaitescu.2017}, or if there is disorder on the values of $t_0$.

\begin{figure}[t!]
\centering
\includegraphics[width=0.9\columnwidth]{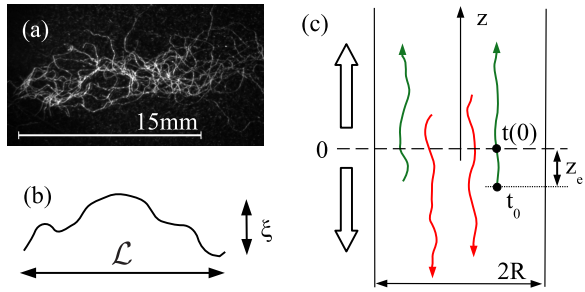}
\caption{(a)Fibers of cotton. (b)Length $\mathcal{L}$ and tortuosity $\xi$ of fiber. (c) Separation of a yarn at a plane $z=0$. Arrows show the directions relatively to the plane $z=0$.}
\label{fig4}
\end{figure}

{\it Staples yarn.} We now apply our results to a yarn made of an assembly of fibers of length $\mathcal{L}$ as shown in fig.\ref{fig4}. Fig.4(c) shows a yarn which separates in two parts from an arbitrary plane $z=0$. A fiber with center located above this plane rises. Let's $z_e$ the distance between the end of the fiber and the plane, and $t_0$ the tension at the end of the fiber. Integrating \eqref{eq5} from $-z_e$ to $0$ gives $t(z=0)=t_0~\exp{( \mu r z_e /p^2)}$. By symmetry, the relation is the same for a descending fiber. Noting $\mathcal{P}(z_e)~dz_e$ the probability that fiber ends at a distance between $z_e$ and $z_e+dz_e$, the total separating force is then:
\begin{subequations}
\begin{align}
T_M&=\int_{r=0}^{r=R} dN(r)
\int_{z_e} t_0~\exp{(\mu r z_e/p^2)}~\mathcal{P}(z_e)dz_e\label{eq7a} \\
&=Nt_0 \frac{2[\exp{(\HT/2)}-(1+(\HT/2))]}{(\HT/2)^2}\label{eq7b}
\end{align}
\end{subequations}
where $\HT= \mu R \LL/p^2$. We used $dN(r)=2 N r dr/R^2$ with $N$ the number of fiber in one section, and assumed an uniform distribution of ends of fibers $\mathcal{P}(z_e)=2/\LL$ for $0\le z_e\le\LL/2$.  The tension $Nt_0$ that the yarn may support without twist is then amplified by a factor $A(\HT)=2[\exp{(\HT/2)}-(1+(\HT/2))]/(\HT/2)^2$. We expect that the exponential amplification still occurs for various distribution $\mathcal{P}(z_e)$: i.e.
taking $\mathcal{P}(z_e)$ as a Dirac distribution $\delta(z_e-L)$ in \eqref{eq7a}, we recover \eqref{eq6b}. Exponential amplification should also occurs in case of disordered values of $t_0$, or if fibers trajectories are not perfectly helicoidal.

{\it Critical Hercules twist Number and Spinning Transition.}
This amplification factor $A(\HT)$ increases nearly exponentially with $\HT$. However, the maximum traction $T_M$ cannot be larger that the force $T_r$ for which the rupture of the fibers occurs. We note $\HT_c$ the critical value of the Hercules twist Number which verifies $T_r=Nt_0A(\HT_c)$. It occurs at a point $C$ on fig.1(e). $\HT_c$ separates weakly twisted yarns ($\HT<\HT_c$) that fail by sliding of fibers, from highly twisted yarns ($\HT >\HT_c$) that fail by breaking of fibers.\\
\indent A typical value of $\HT_c$ for a yarn made of identical fibers of diameter $d$ and of length $\LL$ may be evaluated. Noting $E$ the Young Modulus, and $\varepsilon_r$ the deformation of fibers at rupture, and dropping constant numerical factor, the rupture tension is $t_r\sim \varepsilon_r E d^2$ for a fiber, and $T_r=Nt_r$ for a yarn. Since fibers are slender objects, we take $t_0$ as the force necessary to straighten into a yarn the fibers that are initially bent. Noting $\xi$ the initial flexion of the fibers (fig.4b) we have  $t_0\sim E d^4 \xi/\LL^3$. It follows that $A(\HT_c)=t_r/t_0\sim \varepsilon_r \LL^3 / \xi d^2$.  For cotton fibers with $\LL=30$~mm, $d=16$~$\mu$m, $\mu_m=0.48$~\cite{morton.book,belser.1968}$, \varepsilon_r \simeq  0.08$, and $\xi \sim \LL/3$: $A(\HT_c)\sim 10^5$, and $\HT_c\simeq 33$. The associated pitch for a yarn of radius $R=80$~$\mu$m is $P=2\pi \sqrt{\mu R \LL/\HT_c}\simeq 1.2$~mm. From a microscopic inspection of the yarn, we measured a similar value of the pitch $P\simeq 1.5$~mm. For fibers made of an identical material with $\xi \sim \LL$, and dropping non exponential term in $A(\HT)\sim\exp(\HT/2)$, we obtain the simple scaling $\HT_c \sim 4 \ln(\LL/d)$: $\HT_c$ is in the range $20-40$ when $\LL/d$ varies between $10^2$ to $10^4$.\\

{\it Optimal yarn.}
The maximum resistance of a yarn is attain for $\HT\ge\HT_c$, but it is possible to attain this value? Indeed, twisting a yarn elongates the fibers which may break: twisting too much a yarn reduces its strength, a fact already noticed by Galileo~\cite{galileo}. The elongation may be evaluated: a length $dz$ of an initially straight fiber at $r=R$ becomes $ds= dz \sqrt{1+\gamma^2}$ after the twist of the yarn. The deformation $\varepsilon=(ds-dz)/dz\simeq \gamma^2/2$ should be lower than $\varepsilon_r$, so that the twist must verify $\gamma^2<2\varepsilon_r$. The maximum attainable value of $\HT$ without breaking of fibers is then $\HT_r=2 \mu\varepsilon_r\LL/R$. For a maximal resistance without breaking due to twist we must have $\HT_c\le\HT\le\HT_r$, so that:
\begin{align}R \leq R_{opt} = 2 \mu \varepsilon_r \LL/ \HT_c\end{align}
where we introduced $R_{opt}$ as the value of the yarn radius $R$ which verifies $\HT_r=\HT_c$. $R_{opt}$ is the largest radius of yarn which may reach $\HT_c$ without breaking of fibers. For cotton fibers, with $\HT_c \simeq 30$, we obtain $R_{opt} \simeq 80~\mu$m which is the value of the radius that we measure for our cotton yarn. Thicker simple yarns may be processed, but will not reach their maximal resistance. Making larger yarns with maximal resistance must be done by putting together elementary yarns of radius $R_{opt}$ as it is done in practice~\cite{hearle.1969.book,pan.2014}.

{\it Concluding remarks.}
From our experiments and our statistical model, a relatively simple picture emerges to properly describe the spinning transition of yarn: the twist on the fiber creates a confining potential. The tangential force variations are then proportional to tension, creating exponential decay of the tension. Although the model is very simple, the experimental variations on model yarns are very well captured. This means that a more refined description of the disorder in the fibers arrays, potential deviations from helicoidal structures of fibers, or non-linearity arising from non-small curvature ($r \not \ll p$) are presumably of weak importance.

A crucial result of our study is that the force amplification may be properly described with a single non-dimensional number $\HT$ that we named Hercules twist Number. Although it appears to be a quantity of fundamental interest for the yarn processing, this non-dimensional Number has apparently not be previously defined. This name echoes to the situation of interleaved phone book experiment~\cite{alarcon.2016,taub.2021}. In those studies the authors considered a "Hercules Number" $2\mu M^2 \varepsilon/d$, with $\mu$ the friction coefficient, $M$ the number of pages, $\varepsilon$ the sheet thickness and $d$ the distance of overlap between leaves. Writing $\HT$ as $\mu \theta^2 R/L$, the structure of these two non-dimensional numbers appears similar, but with the noticeable difference that $\theta$ is controlled by the deformation of the yarn, whereas $M$ is fixed. It should be interesting to investigate in details if the assembly of frictional objects with different symmetries, such as packing of non-aligned fibers~\cite{verhille.2017} or twisted sheets~\cite{chopin.2020} show similar exponential force variations. Also, is should be interesting to see if recent results on friction effects on bending of layered structures~\cite{poincloux.2021} may be extended to fibrous structures.

Finally, It should be note that our theory is not only qualitative, but also quantitative since $\HT_c \simeq 30$ corresponds to the twist value for real yarns. The exponential increases of the force amplification factor $A(\HT)$, together with the quadratic dependence with the twist angle $\HT \sim \theta^2$ induces that the spinning process appears in practice as a sharp twist-controlled phase transition.

\end{document}